\documentclass[preprint,letterpaper]{revtex4}
\usepackage{graphicx}
\usepackage{dcolumn}
\usepackage{amsmath}

\begin{document}

\title{Calculation of Hydrodynamic Mass for Atomic Impurities in Helium}

\author{Kevin K. Lehmann}
 \email{lehmann@princeton.edu}
\affiliation{Department of Chemistry, Princeton University, Princeton,
  New Jersey 08544}

\date{\today}

\begin{abstract}
\vspace{4mm}
We present a simple numerical procedure for calculating the
irrotational hydrodynamic flow in a helium solvation structure
around a spherical solute in linear motion through superfluid helium.  
The calculation requires only the radial
helium density around the impurity as input.
From the resulting irrotational flow, the helium contribution to the effective
mass of the solute is calculated. 
The results for alkali cations are compared to recent many-body
Variational Monte Carlo (VMC) calculations by M. Buzzacchi, D.~E. Galli, and L.
Reatto (Phys. Rev. B., {\bf 64} 094512 (2001)). 
The helium contribution to the effective masses calculated by the two methods are
12.9(4.6) versus 9.4 u for Li$^+$, 48.2(5.6) versus 52.1 u for Na$^+$, 69.6(4.8)
versus 70.1 u for K$^+$, and 6.4(8.8) versus 6.8 for Cs$^+$, with the VMC
result listed first (with one $\sigma$ statistical error estimate) and the
hydrodynamic result listed second.  For the cases of Na$^+$ and K$^+$, the
hydrodynamic calculation treated the first helium solvation shell as a rigid solid,
as suggested by the VMC calculations; treating the first solvent layer as part of
the superfluid lead to considerable underestimate, $\approx 50\%$, of the mass
increase in both cases.  In all cases, the agreement of the two results are in
within the error estimate of the VMC calculation, demonstrating the accuracy of
hydrodynamic treatment of helium motion even on the atomic length scale.
\end{abstract}

\pacs{PACS number}
\keywords{Quantum Hydrodynamics, Effective Mass, Liquid Helium, Helium
Nanodroplets}
\maketitle

\bibliographystyle{prsty}


The study of atomic ions as microscopic probes of
superfluidity in helium-4 has a rich history~\cite{Schwarz75}.  
Recent progress on both the injection of neutral metal atoms into bulk liquid
helium~\cite{Takahashi93,Takami96} and the study of doped helium
nanodroplets~\cite{Toennies98} have caused a renewed interest in the dynamical
behavior of solutes in this unique quantum liquid.   In particular, concerning the
rotational degrees of freedom, here have recently been several papers that
try to provide a microscopic explaination for the observed sizable fractional
increase in the rotational moments of inertia of all but the fastest rotors when
dissolved in superfluid
helium~\cite{Blume99,Lee99,Kwon99b,Callegari99b,Kwon00,Lehmann01a}.   
In all of the proposed models, the extra moment of inertia arises from helium kinetic
energy induced by rotation of the molecule, but there is disagreement about the way
this should be calculated, and about the physical description of the helium
motion~\cite{Lehmann01b}.

A closely related problem is that of the change
in the effective translational mass of a solute in superfluid helium, which
also arise from helium kinetic energy induced by the requirement that the helium
solvation structure must `follow' a moving impurity. In the {\it quantum
hydrodynamic} model~\cite{Callegari99b,Lehmann_up}, one calculates the classical
velocity potential that describes the irrotational flow which maintains a
constant helium solvation structure in the frame of a moving impurity.  By a theorem
of Lord Kelvin~\cite{Milne,Callegari99b}, this irrotational flow will provide the
minimum kinetic energy flow that satisfies the equation of
continuity.  
A simple example is the case of a moving spherical ``hole'' in
the liquid, for which the
hydrodynamic flow is a dipole field that carries a kinetic energy equal to
one half of that of the liquid mass displaced by the sphere moving at the velocity of
the hole~\cite{Milne}.   For the general case of an inhomogeneous density,
Barrera and Baym~\cite{Barrera72} have presented a solution to the equation of
continuity, based upon a transformation of dipole solution. 
However, this transformed solution is not irrotational, and if fact has a
continuously varying vorticity, and thus is not appropriate for flow in a superfluid,
where vorticity must be quantized.  

This paper presents a general numerical scheme for finding the irrotational solution
of the equation of continuity given a solvation density around a moving spherical
solute.  The hydrodynamic equation can be solved by separation of variables, and
reduces to one dimensional quadrature of the radial
homogeneous and inhomogeneous equations, allowing the mixed boundary conditions to
be satisfied without need for iteration.  
In the limit of infinitesimal solute velocity, the given solution can be
shown~\cite{Lehmann_up} to provide a variationally optimized helium ground state
wave function, assuming a one-body phase function.  As such, it should provide a
rigorous lower bound on the true increase in effective mass.  

In order to test the quantitative accuracy of this approximation, we have compared
the calculated mass increase for Li$^+$, Na$^+$, K$^+$, and Cs$^+$ with those
calculated by Buzzacchi, Galli, and Reatto~\cite{Buzzacchi01} using a Variational
Monte Carlo (VMC) treatment of the explicit many-body
problem~\cite{Pederiva94,Duminuco00}.  The input to the hydrodynamic calculations
were the helium radial densities calculated by these same authors using the same
method; thus the comparison provides a direct test of the accuracy of the 
one-particle hydrodynamic treatment on atomic length scales. 

\section{Calculation Method}

Consider an atomic solute in superfluid He that has a solvation
structure with radial number density $\rho(r)$.  We assume
that the solute is moving with velocity $V$ (in a direction we take as
the $z$ axis), and that the solvation structure adiabatically moves
with the atom.  This generates an irrotational flow in the
helium of velocity ${\bf v} = - \mbox{\boldmath$\nabla$} \phi$, where
$\phi$ is the velocity potential, which must satisfy the equation of
continuity:
\begin{equation}
\mbox{\boldmath$\nabla$} (\rho \mbox{\boldmath$\nabla$} \phi) = \frac{{\rm
d}\rho}{{\rm d}t}  = - (\mbox{\boldmath$\nabla$} \rho) \cdot ( V \hat{z} )
\end{equation}
If we write $\phi = \tilde{\phi}(r) \, V \cos(\theta)$
where $r$ is the distance from the impurity atom and $\theta$ is the angle with the
$\hat{z}$ axis, we find that the hydrodynamic equation of continuity is solved if
$\tilde{\phi}$ satisfies the following radial equation:
\begin{equation}
\frac{{\rm d}^2 \tilde{\phi}}{{\rm d}r^2} + 
\frac{2}{r} \frac{{\rm d} \tilde{\phi}}{{\rm d}r} -
\frac{2}{r^2}\tilde{\phi} + 
\left( \frac{{\rm d}\ln \rho}{{\rm d}r} \right) 
\left( \frac{{\rm d} \tilde{\phi}}{{\rm d}r} \right) =
-  \left( \frac{{\rm d}\ln \rho}{{\rm d}r} \right) \label{eq:tildaphi}
\end{equation}

At long range from the atom, the density must approach the bulk
value, $\rho_{\rm e}$, and $\tilde{\phi} \rightarrow A/r^2$.  On the inner wall
of the solvation structure,
$r_{\rm i}$ (where the helium density vanishes rapidly),
$\frac{{\rm d}\tilde{\phi}}{{\rm d}r} \rightarrow -1$.  The general solution
to the inhomogeneous equation~\ref{eq:tildaphi} can be written as a sum of any
inhomogeneous solution plus any linear combination of the homogeneous solutions.  The
homogeneous equation is found by setting the right hand side of
Eq.~\ref{eq:tildaphi} equal to zero.   The inhomogeneous solution that satisfies the
boundary conditions was found as follows.  At large $r_{\rm o}$, inhomogeneous and
homogeneous solutions were started assuming the asymptotic form.  The value
of $A = r_{\rm i}^3/2$, which is the correct value for the uniform density
case, was used to start the solutions. The two solutions are numerically
integrated until
$r = r_{\rm i}$, at which point the derivatives of the homogeneous
solution, $\tilde{\phi}_{\rm h}$ and the inhomogeneous solution, $\tilde{\phi}_{\rm
inh}$ are used to determine the constant $B$, equal to:
\begin{equation}
B = - \frac{1 + \left( \frac{{\rm d} \tilde{\phi}_{\rm inh}}{{\rm
d}r}\right)_{r_{\rm i}}}{\left(\frac{{\rm d} \tilde{\phi}_{\rm h}}{{\rm
d}r}\right)_{r_{\rm i}}}
\end{equation}
The inhomogeneous solution satisfying the boundary conditions is:
\begin{equation}
\tilde{\phi}(r) = \tilde{\phi}_{\rm inh}(r) + B \, \tilde{\phi}_{\rm h}(r)
\end{equation}

The hydrodynamic kinetic energy is found by integrating the
helium kinetic energy, which is proportional to $V^2$.  This
allows us to define a hydrodynamic mass, $M_{\rm h}$, by
\begin{equation}
\frac{M_{\rm h}}{M_{\rm He}} = V^{-2} \int \rho \, |\mbox{\boldmath$\nabla$}
\phi|^2 {\rm d}V  \nonumber 
\end{equation}
\begin{equation}
=   \frac{4 \pi}{3} \int \rho(r) 
\left[ \left(\frac{{\rm d} \tilde{\phi}}{{\rm d}r}\right)^2 +
\frac{2 \tilde{\phi}(r)^2}{r^2} \right] r^2 {\rm d}r
\end{equation}
This integral is evaluated from the numerical solution
over the domain $r_{\rm i} \le r \le r_{\rm o}$.  If we assume that
the solution for $r > r_{\rm o}$ is given by the asymptotic form,
then this gives an additional contribution of 
$(8\pi/3) \, M_{\rm He} \, \rho_e \, \tilde{\phi}(r_{\rm o})^2 \, r_{\rm o}$
to the integral defining $M_{\rm h}$.
It has been checked that this definition of $M_{\rm h}$ gives the
correct value of one half the displaced helium mass for
the case of a hole in helium of uniform density.

It is also possible to use the hydrodynamic equation to
transform the integral for the effective mass to give:
\begin{eqnarray}
\frac{M_{\rm h}}{M_{\rm He}} &=&  V^{-2} \left[ 
-\int \phi \left( \frac{{\rm d} \rho}{{\rm d}t}\right) {\rm d}V 
+ \int \rho \, \phi \, \mbox{\boldmath$\nabla$}\phi \cdot {\rm d}S \right]
\nonumber \\
&=&  \frac{4 \pi}{3} \left [ \int_{r_{\rm i}}^{r_{\rm o}}
\tilde{\phi}(r) \left( \frac{{\rm d} \rho}{{\rm d}r}\right) r^2 {\rm d}r
+ \rho(r_{\rm i}) \, \tilde{\phi}(r_{\rm i}) \, r_{\rm i}^2 \right]
\end{eqnarray}
There is no contribution to the volume integral in the region $r > r_{\rm
o}$ because by assumption, $\left( \frac{{\rm d} \rho}{{\rm d}r}\right) =
0$ in this region.
The two estimates for $M_{\rm He}$ need agree only if $\tilde{\phi}$ is a
solution of the continuum hydrodynamic equation, and thus 
a comparison between them
provides a test of the convergence of the numerical solution and the
size of spacing used for integration.

\section{Application to Alkali Cations}

Figure 1 shows the helium solvation densities around Li$^{+}$,
Na$^{+}$, K$^{+}$, and Cs$^{+}$, as calculated by VMC~\cite{Buzzacchi01}
using a trial function of the ``shadow function'' form. 
This technique treats solids, liquids, and solid-liquid mixtures with a single
functional form~cite{Pederiva94,Duminuco00}.  It can be noted that the solvation
density goes almost to zero between the first and second solvent layers for K$^{+}$,
and is highly structured in the case of Na$^{+}$.  The VMC calculations find that
the first solvation layer has a highly solid-like order for the Na$^+$ and K$^{+}$
cases with little to no exchange of these atoms between solvent
layers~\cite{Duminuco00,Buzzacchi01}.  This suggests that the first solvation layer
not be treated as part of the fluid but as a fixed mass that moves rigidly with the
cation, as in the snowball model of Atkins~\cite{Atkins59}.  
In contrast, the VMC find substantial mobility between solvation layers
for Li$^+$ and Cs$^+$, suggesting that in these cases that even the highly 
compacted first solvation layer should be treated as part of the superfluid,
and thus be treated as part of the hydrodynamic flow.

Table \ref{tab:mass} contains comparisons of the effective mass for each cation
calculated using a hydrodynamic treatment and estimated from the VMC
calculations.  It is seen that in all four cases, the agreement of the two
estimates is excellent, being within the VMC statistical error estimate.
For the cases of Na$^{+}$ and K$^{+}$, the first solvation shell, with 10
and 12 helium atoms respectively, was treated as a rigid solid and the hydrodynamic
calculations were begun at the minimum density point between the first and second
solvent shells.  In both these cases, treating the entire density with the
hydrodynamic approach yielded a substantial underestimate for the translational
mass; 21.7 u for Na$^+$ and 36.0 u for K$^+$.  The hydrodynamic calculations 
are vastly less computationally expensive than the many-body
treatment~\cite{Buzzacchi_pc}.

It is useful to compare our present results with that of the widely used model
of the cation ``snowball'' due to Atkins~\cite{Atkins59}.  In this
electrostrictive model, the helium is treated as a continuum dielectric material,
whose density is increased near the cation due to the ion-induced dipole
interaction.  For a radius less than $b$, on the order of 5-6~\AA\ (whose value
depends upon the helium liquid-solid surface tension assumed but is independent
of the specific singly charged cation~\cite{Schwarz75}), the helium is predicted to
form a solid ``snowball'' that moves rigidly with the ion.  This snowball
contributes a mass of $\approx 150$\,u to the ion effective mass~\cite{Schwarz75}. 
In addition, there is a hydrodynamic contribution to the mass, expected to be on the
order of the hard sphere value $m_{\rm HS}= \frac{2\pi}{3}b^3 \rho_e m_{\rm He}
\approx 40$\,u.  Taking into account the increased helium density for $r$ slightly
larger than $b$, using the model used by Barrera and Baym~\cite{Barrera72} ($\rho -
\rho_e = \rho_e \lambda (b/r)^4$ with $\lambda = 0.186$), the above hydrodynamic
treatment predicts the hydrodynamic contribution to the mass to be $0.932 m_{\rm
HS}$, which can be compared to the value $0.97 m_{\rm HS}$ reported by Barrera and
Baym~\cite{Barrera72} for their proposed velocity solution which is not
irrotational.  In agreement with the Kelvin minimum energy principle~\cite{Milne},
their solution is higher in kinetic energy and thus predicts a higher mass. 
Comparision with the both the VMC and hydrodynamic results show that the effective
masses, even for the case of rigid first solvation shells, is considerably less than
those predicted by the snowball model of Atkins.

\section{Conclusions}

The present work demonstrates that the hydrodynamic treatment of the
linear motion of a solute through superfluid helium predicts the solvent
contributions to the effective mass of the impurity in quantitative agreement with
more exact many-body approaches, yet requires only a  trivial additional
computational cost once the solvent density has been calculated.  This applies to
helium density well inside the predicted 5-6~\AA\ radius of the liquid-solid surface
in the ``snowball'' model of Atkins~\cite{Atkins59}.  However, the highly ordered
first solvent layer around some ions must be treated as a solid that rigidly moves
with the ion.  The present results compliment our recent hydrodynamic calculations of
solvent contributions to the moments of inertia of molecules solvated in helium,
which were found to be in good agreement with experiment~\cite{Callegari99b}.  The
present results have been directly compared to higher levels of theory and thus
provide a more critical test of the hydrodynamic model, since uncertainties in the
solute-helium potentials do not enter.  We therefore
demonstrate that the hydrodynamic treatment of superfluid helium motion can be
accurate on the atomic scale, directly refuting recent criticisms of its
use~\cite{Kwon00}.

\begin{acknowledgments}
The author would like to thank the M Buzzacchi, D.E. Galli, and L. Reatto for making
their helium solvation densities and their VMC mass predictions available before
publication.  He would also like to thank Roman Schmied for advice on the manuscript
and for checking the expressions.  This work was supported by the National Science
Foundation and the Airforce Office of Scientific Research.
\end{acknowledgments}

\begin{table}[htb]
\centering
\caption{ Helium contribution to the effective translational mass, in atomic mass
units, for four cations calculated by VMC~\cite{Buzzacchi01} compared to the
hydrodynamic calculation of the same quantity (this work).  For the cases of
Na$^+$ and K$^+$, the hydrodynamic calculation treated the first helium solvation
shell as crystalline, as predicted by the VMC calculations.
\newline}
\begin{tabular}{cdd}\hline\hline
Cation & \textrm{VMC} & \textrm{Hydrodynamic} \\ \hline
Li$^+$ & 12.9\,(4.6)  & 9.4 \\
Na$^+$ & 48.2\,(5.6)  & 52.1 \\
K$^+$  & 69.6\,(4.8)  & 70.1 \\
Cs$^+$ & 6.4\,(8.8)   & 6.8 \\
\hline\hline
  \end{tabular}
  \label{tab:mass}
\end{table}

\begin{figure}[htbp]
\includegraphics[width=\textwidth]{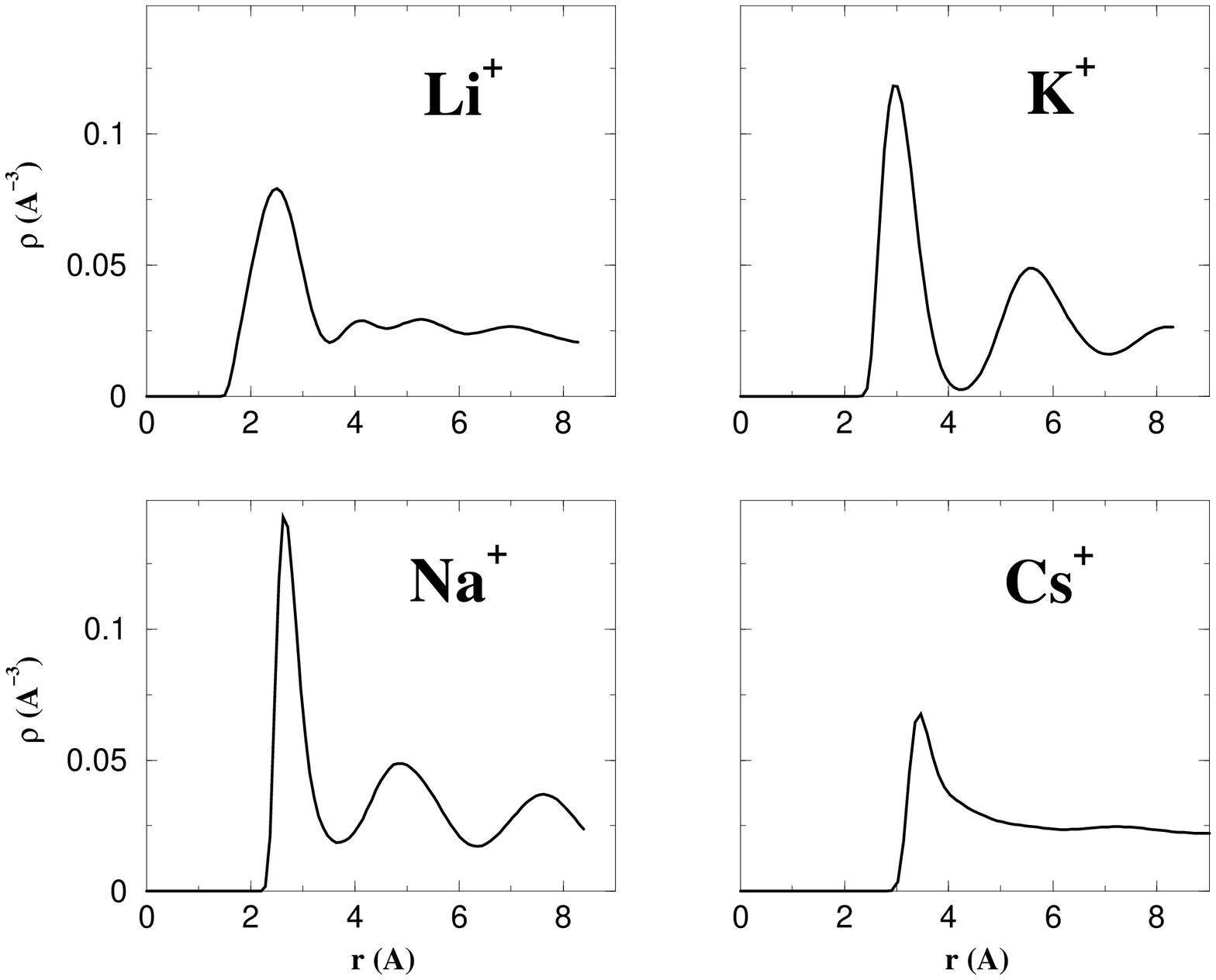}
  \caption{The radial density of $^4$He around the alkali ions. This figure is
reproduced from Ref.~\onlinecite{Buzzacchi01}.}
  \label{fig:exptl}
\end{figure}


\begin{thebibliography}{10}

\bibitem{Schwarz75}
K.W. Schwarz,  in {\em Advances in Chemical Physics}, edited by I. Prigogine
  and S.~A. Rice (John Wiley \& Sons, New York, 1975), Vol.~33, p.\ 1.

\bibitem{Takahashi93}
Y. Takahashi, K. Sano, T. Kinoshita, and T. Yabuzaki, Physical Review Letters
  {\bf 71},  1035  (1993).

\bibitem{Takami96}
M. Takami, Comments on Atomic and Molecular Physics {\bf 32},  219  (1996).

\bibitem{Toennies98}
J.~P. Toennies and A.~F. Vilesov, Annual Reviews of Physical Chemistry {\bf
  49},  1  (1998).

\bibitem{Blume99}
D. Blume, , M. Mladenovic, M. Lewerenz, and K.~B. Whaley, Journal of Chemical
  Physics {\bf 110},  5789  (1999).

\bibitem{Lee99}
E. Lee, D. Farrelly, and K.~B. Whaley, Physical Review Letters {\bf 83},  3812
  (1999).

\bibitem{Kwon99b}
Y. Kwon and K.~B. Whaley, Physical Review Letters {\bf 83},  4108  (1999).

\bibitem{Callegari99b}
C. Callegari, A. Conjusteau, I. Reinhard, K.~K. Lehmann, G. Scoles, and F.
  Dalfovo, Physical Review Letters {\bf 83},  5058  (1999).

\bibitem{Kwon00}
Y. Kwon, P. Huang, M.~V. Patel, D. Blume, and K.~B. Whaley, Journal of Chemical
  Physics {\bf 113},  6469  (2000).

\bibitem{Lehmann01a}
Kevin~K. Lehmann, Rotation in liquid $^4$He: Lessons from a Toy Model, 2000,
  submitted.

\bibitem{Lehmann01b}
See, for example, the exchange between Kevin~K. Lehmann and Mehul~V. Patel, Faraday
Discussions {\bf 118},  33 (2001).

\bibitem{Lehmann_up}
Kevin~K. Lehmann and Carlo Callegari, Relationship of Quantum Hydrodynamic and
  Two Fluid Models for the Effective Moment of Inertia of Molecules in
  Superfluid $^4$He, 2001, manuscript in preparation.

\bibitem{Milne}
L.~M. Milne-Thomson, {\em Theoretical Hydrodynamics}, fifth ed. (Dover, New
  York, 1996).

\bibitem{Barrera72}
R.~G. Barrera and G. Baym, Physical Review A {\bf 6},  1558  (1972).

\bibitem{Buzzacchi01}
M. Buzzacchi, D.~E. Galli, and L. Reatto, Physical Review B {\bf 64}, 094512
(2001).

\bibitem{Pederiva94}
F. Pederiva, A. Ferrante, S. Fantoni, and L. Reatto, Physical Review Letters
  {\bf 72},  2589  (1994).

\bibitem{Duminuco00}
C.~C. Duminuco, D.E. Galli, and L. Reatto, Physica B {\bf 284-288},  109
  (2000).

\bibitem{Atkins59}
K.~R. Atkins, Physical Review {\bf 116},  1339  (1959).

\bibitem{Buzzacchi_pc}
M. Buzzacchi (private communication) reported that each VMC
run  to calculate the mass required approximately four days on a 16 PC Beowulf
cluster.  In contrast, given the helium solvation density (which is much less
expensive to calculate by Monte Carlo methods than the excitation energies), the
hydrodynamic calculation took on the order of one second. 

\end{thebibliography}
\end{document}